\documentclass{svmult}

\usepackage{amsmath}
\usepackage{amsfonts}
\usepackage{dsfont}
\usepackage{graphicx}
\usepackage{makeidx}
\usepackage{multicol}

\newtheorem{exception}{Exception}

\def\ii{\mathrm{i}}
\def\ij{\mathrm{j}}

\def\cs{\operatorname{cs}}
\def\e{\operatorname{e}}
\def\T{\operatorname{T}}

\begin{document}

\title*{Arithmetic quantum chaos of Maass waveforms}
\author{H. Then}
\institute{Abteilung Theoretische Physik,
           Universit\"{a}t Ulm,
           Albert-Einstein-Allee 11,
           89069 Ulm,
           holger.then@physik.uni-ulm.de}

\maketitle
\begin{abstract}
We compute numerically eigenvalues and eigenfunctions of the
quantum Hamiltonian that describes the quantum mechanics of a
point particle moving freely in a particular three-dimensional
hyperbolic space of finite volume and investigate the
distribution of the eigenvalues.
\end{abstract}

\section{Introduction}
The distribution of the eigenvalues of a quantum Hamiltonian is
a central subject that is studied in quantum chaos. There are some
generally accepted conjectures about the nearest-neighbor spacing
distributions of the eigenvalues.
\par
Unless otherwise stated we use the following assumptions:
The quantum mechanical system is desymmetrized with respect to all
its unitary symmetries, and whenever we examine the distribution of
the eigenvalues we regard them on the scale of the mean level spacings.
Moreover, it is generically believed that after desymmetrization a
generic quantum Hamiltonian possesses no degenerate eigenvalues.
\begin{conjecture}[Berry, Tabor \cite{then:BerryTabor1976}] \index{Berry-Tabor conjecture}
If the corresponding classical system is integrable, the eigenvalues
behave like independent random variables and the distribution of the
nearest-neighbor spacings is close to the Poisson distribution, i.e.
there is no level repulsion.
\end{conjecture}
\begin{conjecture}[Bohigas, Giannoni, Schmit \cite{then:BohigasGiannoniSchmit1984,then:BohigasGiannoniSchmit1986}] \label{then:conj:BGS}
If the corresponding classical system is chaotic, the eigenvalues are
distributed like the eigenvalues of hermitian random matrices
\cite{then:Mehta1991}. The corresponding ensembles depend only on the
symmetries of the system:
\begin{itemize}
\item
For chaotic systems \index{Chaotic systems} without time-reversal invariance the distribution
of the eigenvalues should be close to the distribution of the Gaussian
Unitary Ensemble (GUE) \index{GUE} which is characterized by a quadratic level
repulsion.
\item
For chaotic systems with time-reversal invariance and integer spin
the distribution of the eigenvalues should be close to the
distribution of the Gaussian Orthogonal Ensemble (GOE) \index{GOE} which is
characterized by a linear level repulsion.
\item
For chaotic systems with time-reversal invariance and half-integer
spin the distribution of the eigenvalues should be close to the
distribution of the Gaussian Symplectic Ensemble (GSE) \index{GSE} which is
characterized by a quartic level repulsion.
\end{itemize}
\end{conjecture}
These conjectures are very well confirmed by numerical calculations,
but several exceptions are known. Here are two examples:
\begin{exception}
The harmonic oscillator is classically integrable, but its spectrum
is equidistant.
\end{exception}
\begin{exception}
The geodesic motion on surfaces with constant negative curvature
provides a prime example for classical chaos. In some cases, however,
the nearest-neighbor distribution of the eigenvalues of the Laplacian
on these surfaces appears to be Poissonian.
\end{exception}
``A strange arithmetical structure of chaos'' in the case of surfaces of
constant negative curvature that are generated by arithmetic fundamental
groups was discovered by Aurich and Steiner \cite{then:AurichSteiner1988}, see
also Aurich, Bogomolny, and Steiner \cite{then:AurichBogomolnySteiner1991}.
Deviations from the expected GOE-behaviour in the case of a particular
arithmetic surface were numerically observed by Bohigas, Giannoni, and
Schmit \cite{then:BohigasGiannoniSchmit1986} and by Aurich and Steiner
\cite{then:AurichSteiner1989}. Computations coming out in
\cite{then:AurichSteiner1989,then:AurichSteiner1990} showed, however, that the
level statistics on $30$ generic (i.e. non-arithmetic) surfaces were in
nice agreement with the expected random-matrix theory prediction in
accordance with conjecture \ref{then:conj:BGS}. This has led Bogomolny,
Georgeot, Giannoni, and Schmit
\cite{then:BogomolnyGeorgeotGiannoniSchmit1992}, Bolte, Steil, and Steiner
\cite{then:BolteSteilSteiner1992}, and Sarnak \cite{then:Sarnak1995} to introduce
the concept of arithmetic quantum chaos.
\begin{conjecture}[Arithmetic Quantum Chaos] \index{Quantum chaos!arithmetic}
On surfaces of constant negative curvature that are generated by
arithmetic fundamental groups, the distribution of the eigenvalues of the
quantum Hamiltonian are close to the Poisson distribution. Due to level
clustering small spacings occur comparably often.
\end{conjecture}
\par
We compute numerically the eigenvalues and eigenfunctions of the
Laplacian that describes the quantum mechanics of a
point particle moving freely in the non-integrable three-dimensional
hyperbolic space of constant negative curvature generated by the Picard
group. \index{Picard group} The Picard group is arithmetic and we find that our results
are in accordance with the conjecture of arithmetic quantum chaos.
\par
For the definition of an arithmetic group we refer the reader to
\cite{then:Borel1969}.

\section{Preliminaries: The modular group} \index{Modular group}
For simplicity we first introduce the topology and geometry of the
two-dimensional surface of constant negative curvature that is
generated by the modular group \cite{then:Terras1985a}. It will
then be easy to carry over to the three-dimensional space of constant
negative curvature that is generated by the Picard group.
\par
\begin{figure}
\centering
\includegraphics[width=5cm,height=8cm,angle=-90]{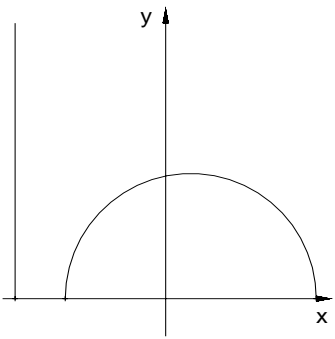}
\caption{Geodesics in the upper half-plane of constant negative
curvature.}
\label{then:fig:then_1}
\end{figure}

The construction begins with the upper half-plane,
\begin{align*}
{\cal H}=\{(x,y)\in\mathds{R}^2; \quad y>0\},
\end{align*}
equipped with the hyperbolic metric of constant negative curvature
\begin{align*}
ds^2=\frac{dx^2+dy^2}{y^2}.
\end{align*}
A free particle on the upper half-plane moves along geodesics, which
are straight lines and semicircles perpendicular to the $x$-axis,
respectively, see figure \ref{then:fig:then_1}.
\par
Expressing a point $(x,y)\in{\cal H}$ as a complex number
$z=x+\ii y$, all isometries of the hyperbolic metric are given by
the group of linear fractional transformations,
\begin{align*}
z\mapsto\gamma z=\frac{az+b}{cz+d}; \quad a,b,c,d\in\mathds{R}, \quad
ad-bc=1,
\end{align*}
which is isomorphic to the group of matrices
\begin{align*}
\gamma=\begin{pmatrix} a&b\\c&d \end{pmatrix}\in
\operatorname{SL}(2,\mathds{R}),
\end{align*}
up to a common sign of the matrix entries,
\begin{align*}
\operatorname{SL}(2,\mathds{R})/\{\pm1\}
=\operatorname{PSL}(2,\mathds{R}).
\end{align*}
\par
\begin{figure}
\centering
\includegraphics[width=5cm,height=8cm,angle=-90]{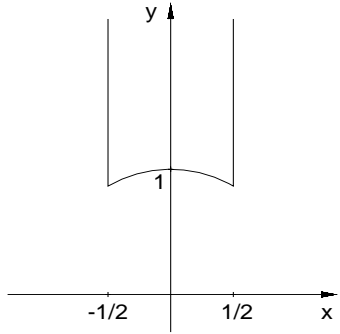}
\caption{The fundamental domain of the modular group.}
\label{then:fig:then_2}
\end{figure}
\begin{figure}
\centering
\includegraphics[width=5cm,height=8cm,angle=-90]{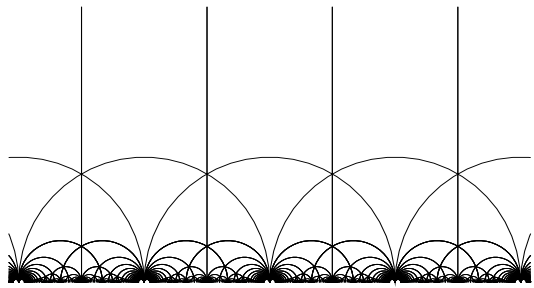}
\caption{The upper half-plane tessellated with isometric copies of
the fundamental domain.}
\label{then:fig:then_3}
\end{figure}

In analogy to the concept of a fundamental cell in a regular lattice
of a crystal we can introduce a fundamental domain of a discrete
group $\Gamma\subset\operatorname{PSL}(2,\mathds{R})$.
\begin{definition}
A fundamental domain of the discrete group $\Gamma$ is an open
subset ${\cal F}\subset{\cal H}$ with the following conditions:
The closure of ${\cal F}$ meets each orbit
$\Gamma z=\{\gamma z;\ \gamma\in\Gamma\}$ at least once,
${\cal F}$ meets each orbit $\Gamma z$ at most once,
and the boundary of ${\cal F}$ has Lebesgue measure zero.
\end{definition}
If we choose the group $\Gamma$ to be the modular group, \index{$\operatorname{PSL}(2,\mathds{Z})$}
\begin{align*}
\Gamma=\operatorname{PSL}(2,\mathds{Z}),
\end{align*}
which is generated by a translation and an inversion,
\begin{align*}
\begin{pmatrix}1&1\\0&1\end{pmatrix}:&\ z\mapsto z+1, \\
\begin{pmatrix}0&-1\\1&0\end{pmatrix}:&\ z\mapsto-z^{-1},
\end{align*}
the fundamental domain of standard shape is
\begin{align*}
{\cal F}=\{z=x+\ii y\in{\cal H}; \quad -\frac{1}{2}<x<\frac{1}{2},
\quad |z|>1\},
\end{align*}
see figure \ref{then:fig:then_2}.
The isometric copies of the fundamental domain
$\gamma{\cal F},\ \gamma\in\Gamma$, tessellate the upper half-plane
completely without any overlap or gap, see figure \ref{then:fig:then_3}.
\par
\begin{figure}
\centering
\includegraphics[width=5cm,height=8cm,angle=-90]{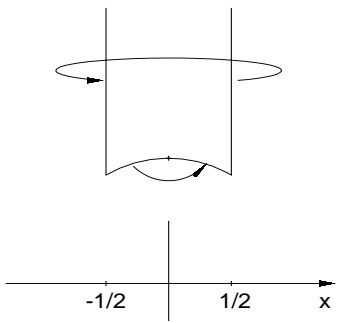}
\caption{Identifying the faces of the fundamental domain according
to the elements of the modular group.}
\label{then:fig:then_4}
\end{figure}

Identifying the fundamental domain ${\cal F}$ and parts of its boundary
with all its isometric copies $\gamma{\cal F},\ \forall\gamma\in\Gamma$,
defines the topology to be the quotient space $\Gamma\backslash{\cal H}$.
The quotient space $\Gamma\backslash{\cal H}$ can also be thought of as
the fundamental domain ${\cal F}$ with its faces glued
according to the elements of the group $\Gamma$, see figure
\ref{then:fig:then_4}.
\par
Any function being defined on the upper half-plane which is
invariant under linear fractional transformations,
\begin{align*}
f(z)=f(\gamma z) \quad \forall\gamma\in\Gamma,
\end{align*}
can be identified with a function living on the quotient space
$\Gamma\backslash{\cal H}$. A function on the quotient space is
tantamount to a function on the fundamental domain with periodic
boundary conditions. Vice versa, any function being defined on the
quotient space can be identified with an automorphic function,
$f(z)=f(\gamma z),\ \forall\gamma\in\Gamma$, living on the upper
half-plane.
\par
With the hyperbolic metric the quotient space $\Gamma\backslash{\cal H}$
inherits the structure of an orbifold. An orbifold locally looks like
a manifold, with the exception that it is allowed to have elliptic
fix-points.
\par
The orbifold of the modular group has one parabolic and two elliptic
fix-points,
\begin{align*}
z=\ii\infty, \quad z=\ii, \quad \text{and} \quad
z=\frac{1}{2}+\ii\frac{\sqrt{3}}{2}.
\end{align*}
The parabolic one fixes a cusp at $z=\ii\infty$ which is invariant
under the parabolic element
\begin{align*}
\begin{pmatrix}1&1\\0&1\end{pmatrix}.
\end{align*}
Hence the orbifold of the modular group is non-compact.
The volume element corresponding to the hyperbolic metric reads
\begin{align*}
d\mu=\frac{dx dy}{y^2},
\end{align*}
such that the volume of the orbifold $\Gamma\backslash{\cal H}$ is finite,
\begin{align*}
\operatorname{vol}(\Gamma\backslash{\cal H})=\frac{\pi}{3}.
\end{align*}
\par
Scaling the units such that $\hbar=1$ and $2m=1$, the stationary
Schr\"{o}dinger equation which describes the quantum mechanics of
a point particle moving freely in the orbifold
$\Gamma\backslash{\cal H}$ becomes
\begin{align*}
(\Delta+\lambda)f(z)=0,
\end{align*}
where the hyperbolic Laplacian is given by
\begin{align*}
\Delta=y^2(\frac{\partial^2}{\partial x^2}
+\frac{\partial^2}{\partial y^2})
\end{align*}
and $\lambda$ is the scaled energy.
We can relate the the eigenvalue problem defined on the orbifold
$\Gamma\backslash{\cal H}$ to the eigenvalue problem defined on the
upper-half space, with the eigenfunctions being subject to the
automorphy condition relative to the discrete group $\Gamma$,
\begin{align*}
f(\gamma z)=f(z) \quad \forall\gamma\in\Gamma.
\end{align*}
In order to avoid solutions that grow exponentially in the cusp, we
impose the boundary condition
\begin{align*}
f(z)=O(y^\kappa) \quad \text{for} \quad z\to\ii\infty
\end{align*}
where $\kappa$ is some positive constant.
\par
The solutions of this eigenvalue problem can be identified with
Maass waveforms \cite{then:Maass1949a}. The identification is worthwhile,
since much is known about Maass waveforms from number theory and
harmonic analysis which will simplify their computation, see e.g.
\cite{then:Selberg1956,then:Roelcke1966,then:Roelcke1967,then:Faddeev1967,then:Shimura1971,then:Hejhal1983,then:Terras1985a,then:Miyake1989,then:Venkov1990,then:Iwaniec1995}.

\section{The \index{Picard group} Picard group}
\begin{figure}
\centering
\includegraphics[width=5cm,height=8cm,angle=-90]{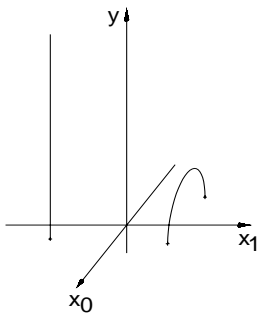}
\caption{Geodesics in the upper half-space of constant negative
curvature.}
\label{then:fig:then_5}
\end{figure}

In the three-dimensional case one considers the upper-half space,
\begin{align*}
{\cal H}=\{(x_0,x_1,y)\in\mathds{R}^3; \quad y>0\}
\end{align*}
equipped with the hyperbolic metric
\begin{align*}
ds^2=\frac{dx_0^2+dx_1^2+dy^2}{y^2}.
\end{align*}
The geodesics of a particle moving freely in the upper half-space
are straight lines and semicircles perpendicular to the
$x_0$-$x_1$-plane, respectively, see figure \ref{then:fig:then_5}.
\par
Expressing any point $(x_0,x_1,y)\in{\cal H}$ as a Hamilton
quaternion, $z=x_0+\ii x_1+\ij y$, with the multiplication defined
by $\ii^2=-1,\ \ij^2=-1,\ \ii\ij+\ij\ii=0$, all motions in the upper
half-space are given by linear fractional transformations
\begin{align*}
z\mapsto\gamma z=(az+b)(cz+d)^{-1}; \quad a,b,c,d\in\mathds{C}, \quad
ad-bc=1.
\end{align*}
The group of these transformations is isomorphic to the group of matrices
\begin{align*}
\gamma=\begin{pmatrix}a&b\\c&d\end{pmatrix}\in
\operatorname{SL}(2,\mathds{C})
\end{align*}
up to a common sign of the matrix entries,
\begin{align*}
\operatorname{SL}(2,\mathds{C})/\{\pm1\}
=\operatorname{PSL}(2,\mathds{C}).
\end{align*}
The motions provided by the elements of $\operatorname{PSL}(2,\mathds{C})$
exhaust all orientation preserving isometries of the hyperbolic metric
on ${\cal H}$.
\begin{remark}
If one wants to avoid using quaternions, the point
$(x_0,x_1,y)\in{\cal H}$ can be expressed by
$(x,y)\in\mathds{C}\times\mathds{R}$ with $x=x_0+\ii x_1$ and $y>0$.
But then the linear fractional transformation look somewhat more
complicated,
\begin{align*}
(x,y)\mapsto\gamma(x,y)=\big(
\frac{(ax+b)(\bar{c}\bar{x}+\bar{d})+a\bar{c}y^2}{|cx+d|^2+|cy|^2},
\frac{y}{|cx+d|^2+|cy|^2}\big).
\end{align*}
In order to keep the notation simple we hence use quaternions.
\end{remark}
\par
\begin{figure}
\centering
\includegraphics[width=5cm,height=8cm,angle=-90]{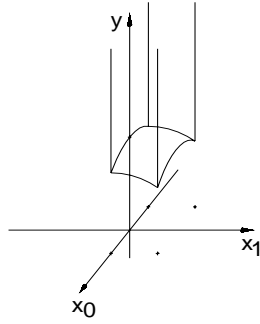}
\caption{The fundamental domain of the Picard group.}
\label{then:fig:then_6}
\end{figure}
\begin{figure}
\centering
\includegraphics[width=5cm,height=8cm,angle=-90]{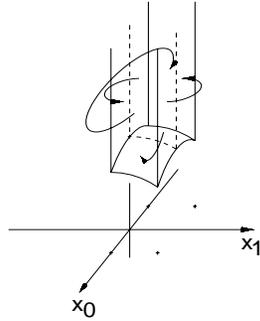}
\caption{Identifying the faces of the fundamental domain according
to the elements of the Picard group.}
\label{then:fig:then_7}
\end{figure}

We now choose the discrete group
$\Gamma\subset\operatorname{PSL}(2,\mathds{C})$ 
generated by the cosets of three elements,
\begin{align*}
\begin{pmatrix}1&1\\0&1\end{pmatrix}, \quad
\begin{pmatrix}1&\ii\\0&1\end{pmatrix}, \quad
\begin{pmatrix}0&-1\\1&0\end{pmatrix},
\end{align*}
which yield two translations and one inversion,
\begin{align*}
z\mapsto z+1, \quad z\mapsto z+\ii, \quad z\mapsto-z^{-1}.
\end{align*}
This group $\Gamma$ is called the Picard group. The three motions
generating $\Gamma$, together with the coset of the element
\begin{align*}
\begin{pmatrix}\ii&0\\0&-\ii\end{pmatrix}
\end{align*}
that is isomorphic to the symmetry
\begin{align*}
z=x+\ij y\mapsto\ii z\ii=-x+\ij y,
\end{align*}
can be used to construct the fundamental domain of standard shape
\begin{align*}
{\cal F}=\{z=x_0+\ii x_1+\ij y\in{\cal H}; \quad
-\frac{1}{2}<x_0<\frac{1}{2}, \quad 0<x_1<\frac{1}{2}, \quad
|z|>1\},
\end{align*}
see figure \ref{then:fig:then_6}.
Identifying the faces of the fundamental domain according to the
elements of the group $\Gamma$ leads to a realization of the
quotient space $\Gamma\backslash{\cal H}$, see figure \ref{then:fig:then_7}.
\par
With the hyperbolic metric the quotient space $\Gamma\backslash{\cal H}$
inherits the structure of an orbifold that has one parabolic and four
elliptic fix-points,
\begin{align*}
z=\ij\infty, \quad z=\ij, \quad z=\frac{1}{2}+\ij\sqrt{\frac{3}{4}},
\quad z=\frac{1}{2}+\ii\frac{1}{2}+\ij\sqrt{\frac{1}{2}}, \quad
z=\ii\frac{1}{2}+\ij\sqrt{\frac{3}{4}}.
\end{align*}
The parabolic fix-point corresponds to a cusp at $z=\ij\infty$ that
is invariant under the parabolic elements
\begin{align*}
\begin{pmatrix}1&1\\0&1\end{pmatrix}, \quad \text{and} \quad
\begin{pmatrix}1&\ii\\0&1\end{pmatrix}.
\end{align*}
The volume element deriving from the hyperbolic metric reads
\begin{align*}
d\mu=\frac{dx_0dx_1dy}{y^3},
\end{align*}
such that the volume of the non-compact orbifold
$\Gamma\backslash{\cal H}$ is finite \cite{then:Humbert1919},
\begin{align*}
\operatorname{vol}(\Gamma\backslash{\cal H})=\frac{\zeta_K(2)}{4\pi^2}
\simeq0.305
\end{align*}
where
\begin{align*}
\zeta_K(s)=\frac{1}{4}\sum_{\nu\in\mathds{Z}[\ii]-\{0\}}
(\nu\bar{\nu})^{-s}, \quad \Re s>1,
\end{align*}
is the Dedekind zeta function.
\par
We are interested in the eigenfunctions of the Laplacian,
\begin{align*}
\Delta=y^2\big(\frac{\partial^2}{\partial x_0^2}+\frac{\partial^2}
{\partial x_1^2}+\frac{\partial^2}{\partial y^2}\big)-y\frac{\partial}
{\partial y},
\end{align*}
which determine the quantum mechanics of a particle moving freely in
the orbifold $\Gamma\backslash{\cal H}$. As in the preceding
section we identify the solutions with \index{Maass waveforms} Maass waveforms
\cite{then:Maass1949b}.
\par
Since the Maass waveforms are automorphic, and therefore periodic in
$x_0$ and $x_1$, it follows that they can be expanded into a Fourier series,
\begin{align}
f(z)=u(y)+\sum_{\beta\in\mathds{Z}[\ii]-\{0\}}
a_{\beta}yK_{\ii r}(2\pi|\beta|y)\e^{2\pi\ii\Re\beta x},
\label{then:FourierExpansion}
\end{align}
where
\begin{align*}
u(y)=\begin{cases} b_0 y^{1+\ii r}+b_1 y^{1-\ii r}&
\text{if $r\not=0$},\\ b_2 y+b_3 y\ln y& \text{if $r=0$}.
\end{cases}
\end{align*}
$K_{\ii r}(x)$ is the K-Bessel function whose order is connected with
the eigenvalue $\lambda$ by
\begin{align*}
\lambda=r^2+1.
\end{align*}
If a Maass waveform vanishes in the cusp,
\begin{align*}
\lim_{z\to\ij\infty}f(z)=0,
\end{align*}
it is called a Maass cusp form. Maass cusp forms are square integrable
over the fundamental domain, $\langle f,f\rangle<\infty$, where
\begin{align*}
\langle f,g\rangle=\int_{\Gamma\backslash{\cal H}}\bar{f}g\,d\mu
\end{align*}
is the Petersson scalar product.
\par
According to the Roelcke-Selberg spectral resolution of the Laplacian
\cite{then:Roelcke1966,then:Roelcke1967}, its spectrum contains both a discrete
and a continuous part. The discrete part is spanned by the constant
eigenfunction $f_0$ and a countable number of Maass cusp forms
$f_1,f_2,f_3,\ldots$ which we take to be ordered with increasing
eigenvalues, $0=\lambda_0<\lambda_1\le\lambda_2\le\lambda_3\le\ldots$.
The continuous part of the spectrum $\lambda\ge1$ is spanned by the
Eisenstein series $E(x,1+\ii r)$ which are known analytically
\cite{then:Kubota1973,then:ElstrodtGrunewaldMennicke1985}. The Fourier coefficients
of the functions $\Lambda_K(1+\ii r)E(x,1+\ii r)$ are given by
\begin{align*}
b_0=\Lambda_K(1+\ii r), \quad b_1=\Lambda_K(1-\ii r), \quad
a_{\beta}=2\sum_{\substack{\lambda,\mu\in\mathds{Z}[\ii] \\
\lambda\mu=\beta}} \big|\frac{\lambda}{\mu}\big|^{\ii r},
\end{align*}
where
\begin{align*}
\Lambda_K(s)=4\pi^{-s}\Gamma(s)\zeta_K(s)
\end{align*}
has an analytic continuation into the complex plane except for a
pole at $s=1$.
\par
Normalizing the Maass cusp forms according to
\begin{align*}
\langle f_n,f_n\rangle=1,
\end{align*}
we can expand any square integrable function
$\phi\in L^2(\Gamma\backslash{\cal H})$ in terms of Maass waveforms,
\cite{then:ElstrodtGrunewaldMennicke1998},
\begin{align*}
\phi(z)=\sum_{n\ge0}\langle f_n,\phi\rangle f_n(z)+\frac{1}{2\pi\ii}
\int_{\Re s=1}\langle E(\cdot,s),\phi\rangle E(z,s)\,ds.
\end{align*}
\par
The eigenvalues and their associated Maass cusp forms are
not known analytically. Thus, one has to approximate them
numerically. Previous calculations of eigenvalues for the Picard
group can be found in \cite{then:SmotrovGolovchanskii1991,then:Huntebrinker1996,
then:GrunewaldHuntebrinker1996,then:Steil1999}. By making use of the Hecke
operators \index{Hecke operator}\cite{then:SmotrovGolovchanskii1991,then:Heitkamp1992} and the
multiplicative relations among the coefficients, Steil
\cite{then:Steil1999} obtained a non-linear system of equations which
allowed him to compute $2545$ consecutive eigenvalues. We extend
these computations with the use of Hejhal's algorithm \cite{then:Hejhal1999}.

\section{Hejhal's algorithm} \index{Hejhal's algorithm}
Hejhal found a linear stable algorithm for computing Maass waveforms
together with their eigenvalues which he used for groups acting
on the two-dimensional hyperbolic plane \cite{then:Hejhal1999}, see also
\cite{then:SelanderStrombergsson2002,then:Avelin2002}. We make use of this
algorithm which is based on the Fourier expansion and the automorphy
condition. We apply it for the Picard group acting on the
three-dimensional hyperbolic space. For the Picard group no small
eigenvalues $0<\lambda=r^2+1<1$ exist \cite{then:Stramm1994}.
Therefore, $r$ is real and the term $u(y)$ in the Fourier expansion
of Maass cusp forms vanishes. \index{cusp forms!Maass} Due to the exponential decay of the
K-Bessel function for large arguments (\ref{then:LargeArgumentK})
and the polynomial bound of the coefficients \cite{then:Maass1949b},
\begin{align*}
a_{\beta}=O(|\beta|), \quad |\beta|\to\infty,
\end{align*}
the absolutely convergent Fourier expansion can be truncated,
\begin{align}
f(z)=\sum_{\substack{\beta\in\mathds{Z}[\ii]-\{0\}\\|\beta|\le M}} a_{\beta}yK_{\ii r}(2\pi|\beta|y)\e^{2\pi\ii\Re\beta x}+[[\varepsilon]], \label{then:TruncatedFourierExpansion} \end{align}
if we bound $y$ from below. Given $\varepsilon>0$, $r$, and $y$, we determine the smallest $M=M(\varepsilon,r,y)$ such that the inequalities
\begin{align*} 2\pi My\ge r \quad \text{and} \quad K_{\ii r}(2\pi My)\le\varepsilon\max_{x}(K_{\ii r}(x)) \end{align*}
hold. Larger $y$ allow smaller $M$. In all truncated terms,
\begin{align*} [[\varepsilon]]=\sum_{\substack{\beta\in\mathds{Z}[\ii]-\{0\}\\|\beta|>M}} a_{\beta}yK_{\ii r}(2\pi|\beta|y)\e^{2\pi\ii\Re\beta x}, \end{align*}
the K-Bessel function decays exponentially in $|\beta|$, and already the K-Bessel function of the first truncated summand is smaller than $\varepsilon$ times most of the K-Bessel functions in the sum of (\ref{then:TruncatedFourierExpansion}). Thus, the error $[[\varepsilon]]$ does at most marginally exceed $\varepsilon$. The reason why $[[\varepsilon]]$ can exceed $\varepsilon$ somewhat is due to the possibility that the summands in (\ref{then:TruncatedFourierExpansion}) cancel each other, or that the coefficients in the truncated terms are larger than in (\ref{then:TruncatedFourierExpansion}). By a finite two-dimensional Fourier transformation the Fourier expansion (\ref{then:TruncatedFourierExpansion}) is solved for its coefficients
\begin{align} a_{\gamma}yK_{\ii r}(2\pi|\gamma|y)=\frac{1}{(2Q)^2}\sum_{x\in\mathds{X}[\ii]} f(x+\ij y)\e^{-2\pi\ii\Re\gamma x}+[[\varepsilon]], \label{then:FourierTransformed} \end{align}
where $\mathds{X}[\ii]$ is a two-dimensional equally distributed set of $(2Q)^2$ numbers,
\begin{align*} \mathds{X}[\ii]=\{\frac{k_0+\ii k_1}{2Q}; \quad k_i=-Q+\tfrac{1}{2},-Q+\tfrac{3}{2},\ldots,Q-\tfrac{3}{2},Q-\tfrac{1}{2}, \quad i=0,1\}, \end{align*}
with $2Q>M+|\gamma|$. \par
By automorphy we have
\begin{align*} f(z)=f(z^*), \end{align*}
where $z^*$ is the $\Gamma$-pullback of the point $z$ into the fundamental domain ${\cal F}$,
\begin{align*} z^*=\gamma z, \quad \gamma\in\Gamma, \quad z^*\in{\cal F}. \end{align*}
Thus, a Maass cusp form can be approximated by
\begin{align} f(x+\ij y)=f(x^*+\ij y^*)=\sum_{\substack{\beta\in\mathds{Z}[\ii]-\{0\}\\|\beta|\le M_0}} a_{\beta}y^*K_{\ii r}(2\pi|\beta|y^*)\e^{2\pi\ii\Re\beta x^*}+[[\varepsilon]], \label{then:Pullback} \end{align}
where $y^*$ is always larger or equal than the height $y_0$ of the lowest points of the fundamental domain ${\cal F}$,
\begin{align*} y_0=\min_{z\in{\cal F}}(y)=\frac{1}{\sqrt{2}}, \end{align*}
allowing us to replace $M(\varepsilon,r,y)$ by $M_0=M(\varepsilon,r,y_0)$. \par
Choosing $y$ smaller than $y_0$ the $\Gamma$-pullback $z\mapsto z^*$ of any point into the fundamental domain ${\cal F}$ makes at least once use of the inversion $z\mapsto-z^{-1}$, possibly together with the translations $z\mapsto z+1$ and $z\mapsto z+\ii$. This is called implicit automorphy, since it guarantees the invariance $f(z)=f(-z^{-1})$. The conditions $f(z)=f(z+1)$ and $f(z)=f(z+\ii)$ are automatically satisfied due to the Fourier expansion. \par
Making use of the implicit automorphy by replacing $f(x+\ij y)$ in (\ref{then:FourierTransformed}) with the right-hand side of (\ref{then:Pullback}) gives
\begin{align} a_{\gamma}yK_{\ii r}(2\pi|\gamma|y)=\frac{1}{(2Q)^2}\sum_{x\in\mathds{X}[\ii]}\sum_{\substack{\beta\in\mathds{Z}[\ii]-\{0\}\\|\beta|\le M_0}} a_{\beta}y^*K_{\ii r}(2\pi|\beta|y^*)\e^{2\pi\ii\Re\beta x^*}\e^{-2\pi\ii\Re\gamma x}+[[2\varepsilon]], \label{then:CentralIdentity} \end{align}
which is the central identity in the algorithm. \par
\begin{figure}[t]
\centering
\includegraphics[width=3.75cm,height=6cm,angle=-90]{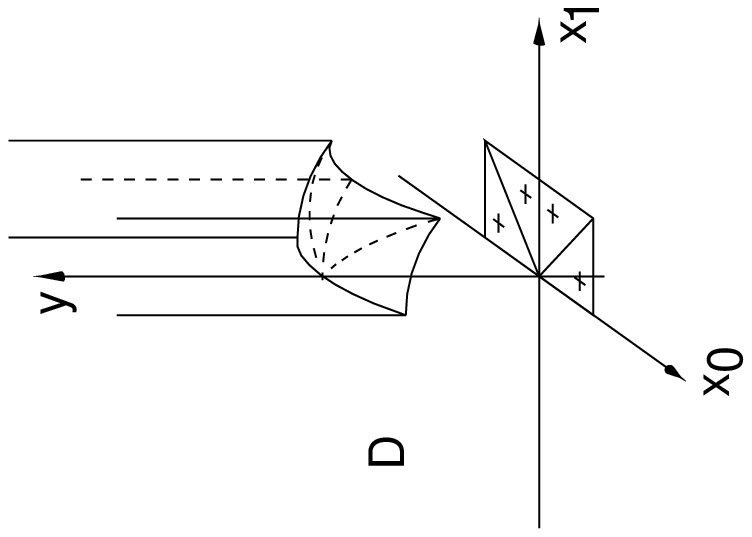} \hspace{-2cm}
\includegraphics[width=3.75cm,height=6cm,angle=-90]{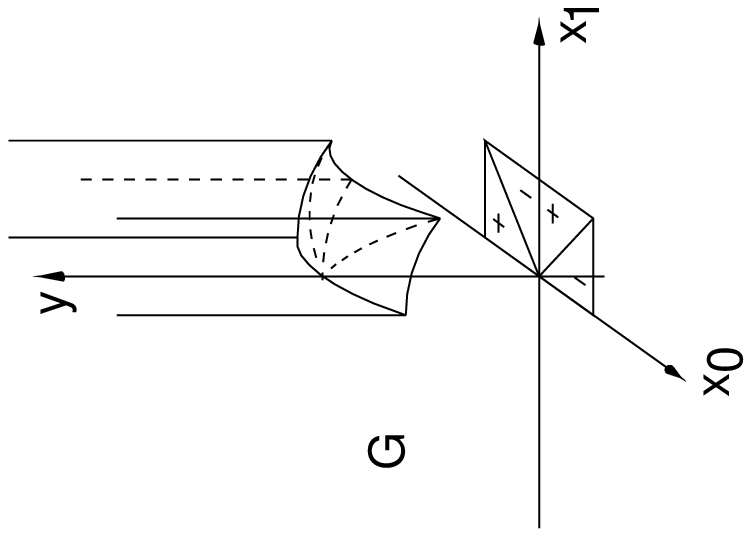} \\
\includegraphics[width=3.75cm,height=6cm,angle=-90]{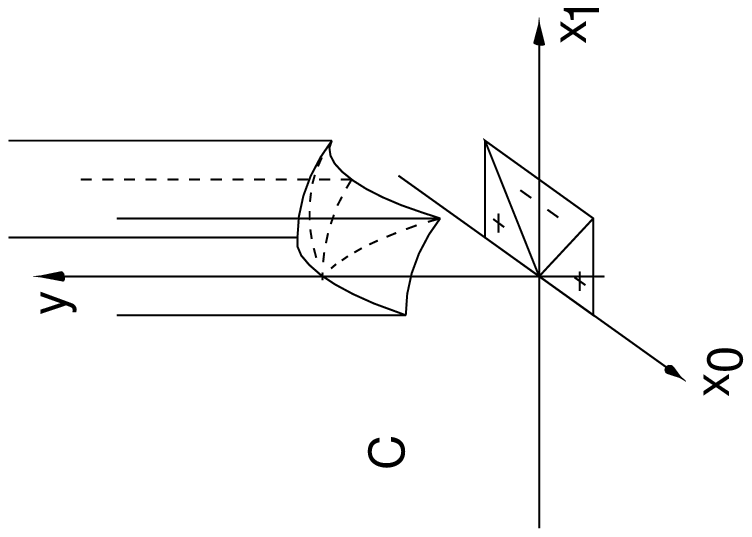} \hspace{-2cm}
\includegraphics[width=3.75cm,height=6cm,angle=-90]{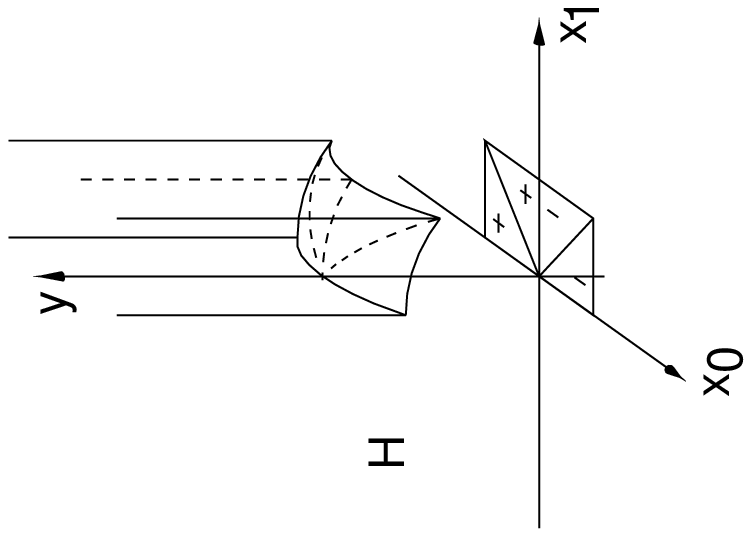}
\caption{The symmetries ${\mathbf D}$, ${\mathbf G}$, ${\mathbf C}$, and ${\mathbf H}$ from top left to bottom right.} \label{then:Symmetries} \end{figure}

The symmetry in the Picard group and the symmetries of the fundamental domain imply that the Maass waveforms fall into four symmetry classes \cite{then:Steil1999} named ${\mathbf D}$, ${\mathbf G}$, ${\mathbf C}$, and ${\mathbf H}$, satisfying
\begin{align*} &{\mathbf D}: \quad f(x+\ij y)=f(\ii x+\ij y)=f(-\bar{x}+\ij y),\\ &{\mathbf G}: \quad f(x+\ij y)=f(\ii x+\ij y)=-f(-\bar{x}+\ij y),\\ &{\mathbf C}: \quad f(x+\ij y)=-f(\ii x+\ij y)=f(-\bar{x}+\ij y),\\ &{\mathbf H}: \quad f(x+\ij y)=-f(\ii x+\ij y)=-f(-\bar{x}+\ij y), \end{align*}
respectively, see figure \ref{then:Symmetries}, from which the symmetry relations among the coefficients follow,
\begin{align*} &{\mathbf D}: \quad a_{\beta}=a_{\ii\beta}=a_{\bar{\beta}},\\ &{\mathbf G}: \quad a_{\beta}=a_{\ii\beta}=-a_{\bar{\beta}},\\ &{\mathbf C}: \quad a_{\beta}=-a_{\ii\beta}=a_{\bar{\beta}},\\ &{\mathbf H}: \quad a_{\beta}=-a_{\ii\beta}=-a_{\bar{\beta}}. \end{align*}
Defining
\begin{align*} \cs(\beta,x)=\sum_{\sigma\in\mathds{S}_{\beta}} s_{\sigma\beta}\e^{2\pi\ii\Re\sigma x}, \end{align*}
where $s_{\sigma\beta}$ is given by
\begin{align*} a_{\sigma}=s_{\sigma\beta}a_{\beta} \end{align*}
and
\begin{align*} \sigma\in\mathds{S}_{\beta}=\begin{cases} \{\beta,\ii\beta,-\beta,-\ii\beta,\bar{\beta},\ii\bar{\beta},-\bar{\beta},-\ii\bar{\beta}\}& \text{if $\bar{\beta}\not\in\{\beta,\ii\beta,-\beta,-\ii\beta\}$},\\ \{\beta,\ii\beta,-\beta,-\ii\beta\}& \text{else}, \end{cases} \end{align*}
the Fourier expansion (\ref{then:FourierExpansion}) of the Maass waveforms can be written
\begin{align*} f(z)=u(y)+\sum_{\beta\in\tilde{\mathds{Z}}[\ii]-\{0\}} a_{\beta}yK_{\ii r}(2\pi|\beta|y)\cs(\beta,x), \end{align*}
where the tilde operator on a set of numbers is defined such that
\begin{align*} \tilde{\mathds{X}}\subset\mathds{X}, \quad \bigcup_{x\in\tilde{\mathds{X}}} \mathds{S}_{x}=\mathds{X}, \quad \text{and} \quad \bigcap_{x\in\tilde{\mathds{X}}} \mathds{S}_{x}=\emptyset \end{align*}
holds. \par
Forgetting about the error $[[2\varepsilon]]$ the set of equations (\ref{then:CentralIdentity}) can be written as
\begin{align} \sum_{\substack{\beta\in\tilde{\mathds{Z}}[\ii]-\{0\}\\|\beta|\le M_0}} V_{\gamma\beta}(r,y)a_{\beta}=0, \quad \gamma\in\tilde{\mathds{Z}}[\ii]-\{0\}, \quad |\gamma|\le M_0, \label{then:HomogeneousSystem} \end{align}
where the matrix $V=(V_{\gamma\beta})$ is given by
\begin{align*} V_{\gamma\beta}(r,y)=\#\{\sigma\in\mathds{S}_{\gamma}\}yK_{\ii r}(2\pi|\gamma|y)\delta_{\gamma\beta}-\frac{1}{(2Q)^2}\sum_{x\in\mathds{X}[\ii]} y^*K_{\ii r}(2\pi|\beta|y^*)\cs(\beta,x^*)\cs(\gamma,-x). \end{align*}
Since $y<y_0$ can always be chosen such that $K_{\ii r}(2\pi|\gamma|y)$ is not too small, the diagonal terms in the matrix $V$ do not vanish for large $|\gamma|$ and the matrix is well conditioned. \par
We are now looking for the non-trivial solutions of (\ref{then:HomogeneousSystem}) for $1\le|\gamma|\le M_0$ that simultaneously give the eigenvalues $\lambda=r^2+1$ and the coefficients $a_{\beta}$. Trivial solutions are avoided by setting one of the coefficients equal to one, $a_{\alpha}=1$. Here we choose $\alpha$ to be $1$, $2+\ii$, $1$, and $1+\ii$, for the symmetry classes ${\mathbf D}$, ${\mathbf G}$, ${\mathbf C}$, and ${\mathbf H}$, respectively. \par
Since the eigenvalues are unknown we discretize the $r$ axis and solve for each $r$ value on this grid the inhomogeneous system of equations
\begin{align} \sum_{\substack{\beta\in\tilde{\mathds{Z}}[\ii]-\{0,\alpha\}\\|\beta|\le M_0}} V_{\gamma\beta}(r,y^{\#1})a_{\beta}=-V_{\gamma\alpha}(r,y^{\#1}), \quad 1\le|\gamma|\le M_0, \label{then:InhomogeneousSystem} \end{align}
where $y^{\#1}<y_0$ is chosen such that $K_{\ii r}(2\pi|\gamma|y^{\#1})$ is not too small for $1\le|\gamma|\le M_0$. A good value to try for $y^{\#1}$ is given by $2\pi M_0y^{\#1}=r$. \par
It is important to check whether
\begin{align*} g_{\gamma}=\sum_{\substack{\beta\in\tilde{\mathds{Z}}[\ii]-\{0\}\\|\beta|\le M_0}} V_{\gamma\beta}(r,y^{\#2})a_{\beta}, \quad 1\le|\gamma|\le M_0, \end{align*}
vanishes where $y^{\#2}$ is another $y$ value independent of $y^{\#1}$. Only if all $g_{\gamma}$ vanish simultaneously the solution of (\ref{then:InhomogeneousSystem}) is independent of $y$. In this case $\lambda=r^2+1$ is an eigenvalue and the $a_{\beta}$'s are the coefficients of the Fourier expansion of the corresponding Maass cusp form. \par
The probability to find an $r$ value such that all $g_{\gamma}$ vanish simultaneously is zero, because the discrete eigenvalues are of measure zero in the real numbers. Therefore, we make use of the intermediate value theorem where we look for simultaneous sign changes in $g_{\gamma}$. Once we have found them in at least half of the $g_{\gamma}$'s we have found an interval which contains an eigenvalue with high probability. By some bisection and interpolation we can see if this interval really contains an eigenvalue, and by nesting up the interval until its size tends to zero we obtain the eigenvalue. \par
In order not to miss eigenvalues which lie close together nor to waste CPU time with a too fine grid, we use the adaptive $r$ grid introduced in \cite{then:Then2002}.

\section{Eigenvalues for the Picard group} We have found $13950$ eigenvalues of the Laplacian for the Picard group in the interval $1<\lambda=r^2+1\le19601$. $4115$ of them belong to eigenfunctions of the symmetry class ${\mathbf D}$, $2805$ to ${\mathbf G}$, $3715$ to ${\mathbf C}$, and $3315$ to ${\mathbf H}$. The smallest eigenvalue is $\lambda=r^2+1$ with $r=6.6221193402528$ which is in agreement with the lower bound $\lambda>\frac{2\pi^2}{3}$ \cite{then:Stramm1994}. Table \ref{then:FirstFewEigenvalues} shows the first few eigenvalues of each symmetry class. They agree with those of Steil \cite{then:Steil1999} up to five decimal places.
\begin{table} \caption{The first few eigenvalues of the Laplacian for the Picard group. Listed is $r$. related to the eigenvalues via $\lambda=r^2+1$.} \label{then:FirstFewEigenvalues} \begin{align*} {\mathbf D}& & {\mathbf G}& & {\mathbf C}& & {\mathbf H}& \\ \\
\ 8&.55525104 & & & \ 6&.62211934 \\
11&.10856737 & & & 10&.18079978 \\
12&.86991062 & & & 12&.11527484 & 12&.11527484 \\
14&.07966049 & & & 12&.87936900 \\
15&.34827764 & & & 14&.14833073 \\
15&.89184204 & & & 14&.95244267 & 14&.95244267 \\
17&.33640443 & & & 16&.20759420 \\
17&.45131992 & 17&.45131992 & 16&.99496892 & 16&.99496892 \\
17&.77664065 & & & 17&.86305643 & 17&.86305643 \\
19&.06739052 & & & 18&.24391070 \\
19&.22290266 & & & 18&.83298996 \\
19&.41119126 & & & 19&.43054310 & 19&.43054310 \\
20&.00754583 & & & 20&.30030720 & 20&.30030720 \\
20&.70798880 & 20&.70798880 & 20&.60686743 \\
20&.81526852 & & & 21&.37966055 & 21&.37966055 \\
21&.42887079 & & & 21&.44245892 \\
22&.12230276 & & & 21&.83248972 & 21&.83248972 \\
22&.63055256 & & & 22&.58475297 & 22&.58475297 \\
22&.96230105 & 22&.96230105 & 22&.85429195 \\
23&.49617692 & & & 23&.49768305 & 23&.49768305 \\
23&.52784503 & & & 23&.84275866 \\
23&.88978413 & 23&.88978413 & 23&.89515755 & 23&.89515755 \\
24&.34601664 & & & 24&.42133829 & 24&.42133829 \\
24&.57501426 & & & 25&.03278076 & 25&.03278076 \\
24&.70045917 & & & 25&.42905483 \\
25&.47067539 & & & 25&.77588591 & 25&.77588591 \\
25&.50724616 & & & 26&.03903968 \\
25&.72392169 & 25&.72392169 & 26&.12361823 & 26&.12361823 \\
\end{align*} \end{table}
We next regard the statistics of the eigenvalues. First, we compare the output of our algorithm with Weyl's law and higher order corrections drawn from \cite{then:Matthies1995}. This serves as a check whether we have found all eigenvalues. We then find it necessary to correct one of the terms in \cite{then:Matthies1995} numerically. Finally, we regard the spectral fluctuations and find that the nearest-neighbor spacing distribution closely resembles that of a Poisson random process as predicted by \cite{then:BogomolnyGeorgeotGiannoniSchmit1992,then:BolteSteilSteiner1992,then:Sarnak1995} and previously observed by \cite{then:Steil1999}. \par
\begin{figure} \centering \includegraphics[width=3.4cm,height=11.7cm,angle=-90]{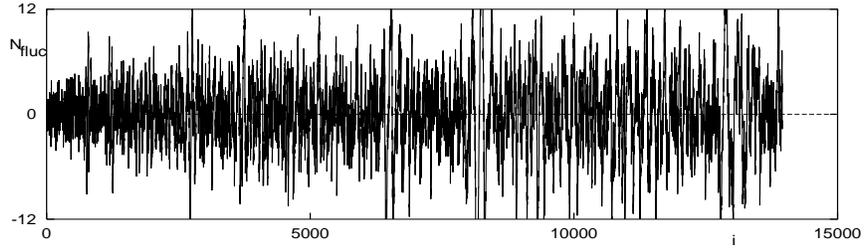} \caption{$N_{fluc}(r_i)$ as a function of $i$ fluctuating around zero.} \label{then:FluctuationsPlot} \end{figure}
\begin{figure} \centering \includegraphics[width=3.4cm,height=11.7cm,angle=-90]{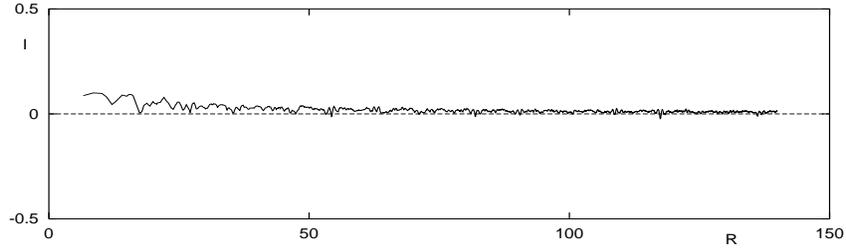} \caption{$I$ as a function of $R$ showing that $I\xrightarrow{R\to\infty}0$.} \label{then:IntegratedFluctuationsPlot} \end{figure}
\begin{figure} \centering \includegraphics[width=6.8cm,height=11.7cm,angle=-90]{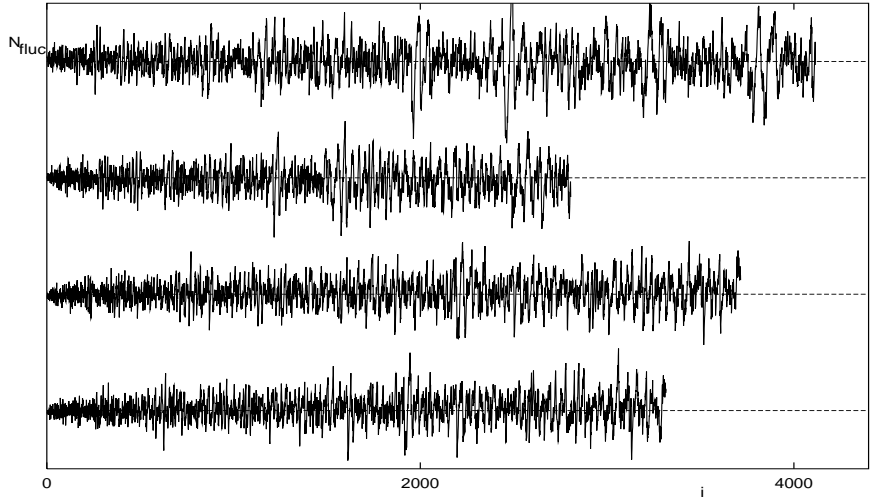} \caption{$N_{fluc}(r_i)$ as a function of $i$ for each symmetry class. The symmetry classes are ${\mathbf D}$, ${\mathbf G}$, ${\mathbf C}$, ${\mathbf H}$ from top to bottom.} \label{then:DeviationsPlot} \end{figure}
\begin{figure} \centering \includegraphics[width=3.4cm,height=11.7cm,angle=-90]{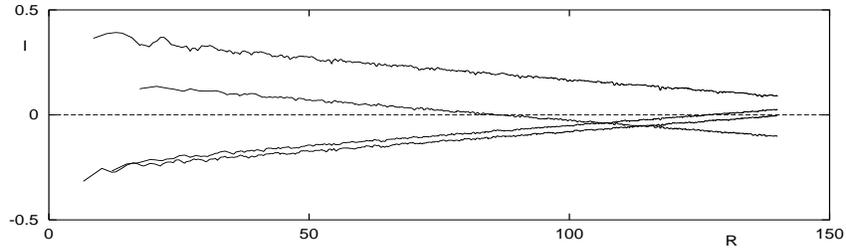} \caption{$I$ as a function of $R$ showing the systematic deviations from $I\xrightarrow{R\to\infty}0$. Each curve belongs to one of the symmetry classes ${\mathbf D}$, ${\mathbf G}$, ${\mathbf C}$, ${\mathbf H}$.} \label{then:IntegratedDeviationsPlot} \end{figure}
\begin{figure} \centering \includegraphics[width=3.4cm,height=11.7cm,angle=-90]{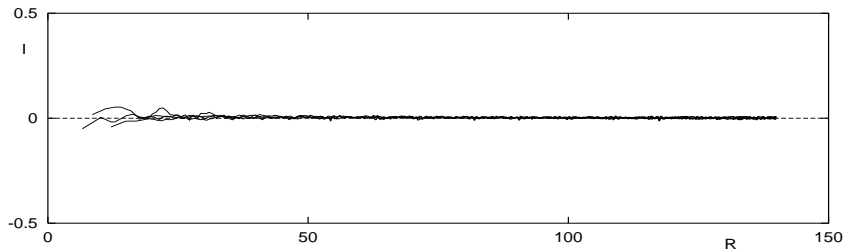} \caption{$I$ as a function of $R$ with the corrected constants. Each curve belongs to one of the symmetry classes. The curves are quite indistinguishable from 0.} \label{then:CorrectedPlot} \end{figure}

In the first step we consider the level counting function
\begin{align*} N(r)=\#\{\,i\ |\ r_i\le r\} \end{align*}
and split it into two parts
\begin{align*} N(r)=\bar{N}(r)+N_{fluc}(r). \end{align*}
Here $\bar{N}$ is a smooth function describing the average increase in the number of levels and $N_{fluc}$ describes the fluctuations around the mean such that
\begin{align*} \lim_{R\to\infty}\frac{1}{R}\int_{1}^{R}N_{fluc}(r)dr=0. \end{align*}
The average increase in the number of levels is given by Weyl's law \cite{then:Weyl1912,then:Avakumovic1956} and higher order corrections have been calculated by Matthies \cite{then:Matthies1995}. She obtained
\begin{align} \bar{N}(r)=\tfrac{\operatorname{vol}({\cal F})}{6\pi^2}r^3+a_2 r \log r+a_3 r+a_4 \label{then:WeylMatthies} \end{align}
with the constants
\begin{align*} a_2&=-\tfrac{3}{2\pi}, \\ a_3&=\tfrac{1}{\pi}[\tfrac{13}{16}\log 2+\tfrac{7}{4}\log\pi-\log\Gamma(\tfrac{1}{4})+\tfrac{2}{9}\log(2+\sqrt{3})+\tfrac{3}{2}], \\ a_4&=-\tfrac{1}{2}. \end{align*}
We compare our results for $N(r)$ with (\ref{then:WeylMatthies}) by defining
\begin{align} N_{fluc}(r)=N(r)-\bar{N}(r). \label{then:Fluctuations} \end{align}
$N_{fluc}$ fluctuates around zero or a negative integer whose absolute value gives the number of missing eigenvalues, see figure \ref{then:FluctuationsPlot}. Unfortunately, our algorithm does not find all eigenvalues in one single run. In the first run it finds about $97\%$ of the eigenvalues. Apart from very few exceptions the remaining eigenvalues are found in the third run. To be more specific, we plotted $N_{fluc}$ decreased by $\frac{1}{2}$, because $N(r)-\bar{N}(r)$ is approximately $\frac{1}{2}$ whenever $\lambda=r^2+1$ is an eigenvalue. Furthermore, we took the eigenvalue $\lambda=0$ into account. We remark that we never find more eigenvalues than predicted by (\ref{then:WeylMatthies}). A plot indicating that $N_{fluc}$ fluctuates around zero is shown in figure \ref{then:IntegratedFluctuationsPlot} where we plotted the integral
\begin{align} I(R)=\frac{1}{R}\int_{1}^{R}N_{fluc}(r)dr. \label{then:IntegratedFluctuations} \end{align}
So far, everything seems to be consistent. Taking the desymmetrized spectra into account (\ref{then:WeylMatthies}) is modified \cite{then:Matthies1995}
\begin{align} \bar{N}(r)=\tfrac{\operatorname{vol}({\cal F})}{24\pi^2}r^3+b_1 r^2+b_2 r \log r+b_3 r+b_4 \label{then:WeylMatthiesDes} \end{align}
with the constants depending on the symmetry class. For the symmetry class ${\mathbf D}$ the constants are given in \cite{then:Matthies1995} as
\begin{align*} b_1&=\tfrac{1}{24}, \\ b_2&=-\tfrac{13}{8\pi}, \\ b_3&=\tfrac{1}{4\pi}[-\tfrac{11}{16}\log 2+\tfrac{19}{4}\log\pi-\log\Gamma(\tfrac{1}{4}) \\ & \quad\quad +\tfrac{2}{9}\log(2+\sqrt{3})+\tfrac{1}{4}\log(3+2\sqrt{2})+\tfrac{13}{2}], \\ b_4&=-\tfrac{47}{72}. \end{align*}
For ${\mathbf G}$
\begin{align*} b_1&=-\tfrac{1}{24}, \\ b_2&=\tfrac{3}{8\pi}, \\ b_3&=\tfrac{1}{4\pi}[\tfrac{37}{16}\log 2+\tfrac{3}{4}\log\pi-\log\Gamma(\tfrac{1}{4}) \\ & \quad\quad +\tfrac{2}{9}\log(2+\sqrt{3})+\tfrac{1}{4}\log(3+2\sqrt{2})-\tfrac{3}{2}], \\ b_4&=-\tfrac{25}{72}. \end{align*}
For ${\mathbf C}$
\begin{align*} b_1&=\tfrac{1}{96}, \\ b_2&=-\tfrac{1}{8\pi}, \\ b_3&=\tfrac{1}{4\pi}[\tfrac{5}{16}\log 2+\tfrac{3}{4}\log\pi-\log\Gamma(\tfrac{1}{4}) \\ & \quad\quad +\tfrac{2}{9}\log(2+\sqrt{3})-\tfrac{1}{4}\log(3+2\sqrt{2})+\tfrac{1}{2}], \\ b_4&=\tfrac{125}{576}. \end{align*}
And for ${\mathbf H}$
\begin{align*} b_1&=-\tfrac{1}{96}, \\ b_2&=-\tfrac{1}{8\pi}, \\ b_3&=\tfrac{1}{4\pi}[\tfrac{21}{16}\log 2+\tfrac{3}{4}\log\pi-\log\Gamma(\tfrac{1}{4}) \\ & \quad\quad +\tfrac{2}{9}\log(2+\sqrt{3})-\tfrac{1}{4}\log(3+2\sqrt{2})+\tfrac{1}{2}], \\ b_4&=\tfrac{163}{576}. \end{align*}
Let $\{r_i\}$ be a sequence related to the consecutive eigenvalues $\lambda=r^2+1$. If we plot $N_{fluc}(r_i)$ as a function of $i$ for the desymmetrized spectra we obtain small deviations which can hardly be seen in figure \ref{then:DeviationsPlot}. But if we plot the integral (\ref{then:IntegratedFluctuations}) we see that $N_{fluc}$ does not really fluctuate around zero. Instead, in figure \ref{then:IntegratedDeviationsPlot} we see systematic deviations,
but the discrepancy is much less than one eigenvalue for each symmetry class. Since the number of eigenvalues is integer-valued we do not assume that we have found less or too many eigenvalues. Therefore, we fit the constants $b_1,b_2,b_3,b_4$ in (\ref{then:WeylMatthiesDes}) and obtain new constants for each of the symmetry classes. Since the integrals $I(R)$ in figure \ref{then:IntegratedDeviationsPlot} show a linear behavior, the constants $b_1$ and $b_2$ seem to be correct. We thus only change the constants $b_3$ and $b_4$ by fitting them numerically. For the symmetry class ${\mathbf D}$ the new constants are
\begin{align*} b_3&= 0.8639... &&\text{ instead of } b_3=0.8679... , \\ b_4&= -0.288... &&\text{ instead of } b_4=-0.653...\,. \end{align*}
For ${\mathbf G}$
\begin{align*} b_3&= 0.0285... &&\text{ instead of } b_3=0.0324... , \\ b_4&= -0.184... &&\text{ instead of } b_4=-0.347...\ . \end{align*}
For ${\mathbf C}$
\begin{align*} b_3&= 0.0150... &&\text{ instead of } b_3=0.0111... , \\ b_4&= -0.062... &&\text{ instead of } b_4=0.217...\,. \end{align*}
And ${\mathbf H}$
\begin{align*} b_3&= 0.0702... &&\text{ instead of } b_3=0.0662... , \\ b_4&= 0.034... &&\text{ instead of } b_4=0.283...\ . \end{align*}
In figure \ref{then:CorrectedPlot} we present the integral (\ref{then:IntegratedFluctuations}) with the corrected constants.
 \par
\begin{figure} \centering \includegraphics[width=6.8cm,height=11.7cm,angle=-90]{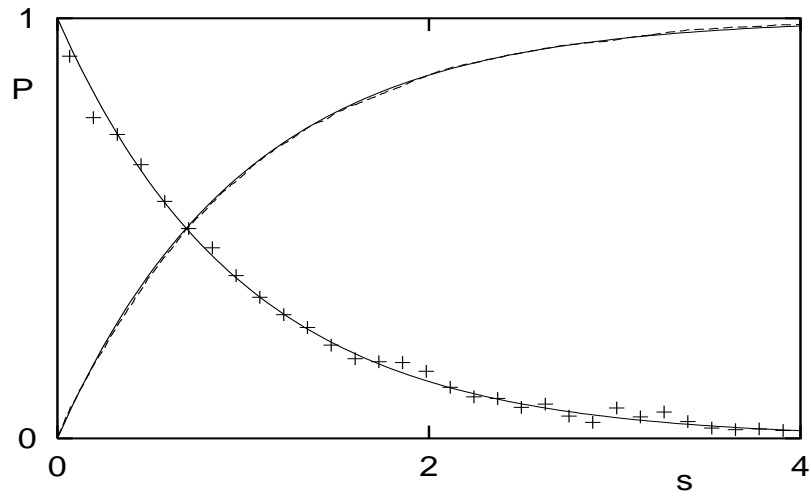} \caption{Level spacing distribution for the symmetry class ${\mathbf D}$. The abscissa displays the spacings $s$. The dashed curve starting at the origin is the integrated distribution. For comparison, the full curves show a Poisson distribution.} \label{then:LevelSpacingsPlotD} \end{figure}
\begin{figure} \centering \includegraphics[width=6.8cm,height=11.7cm,angle=-90]{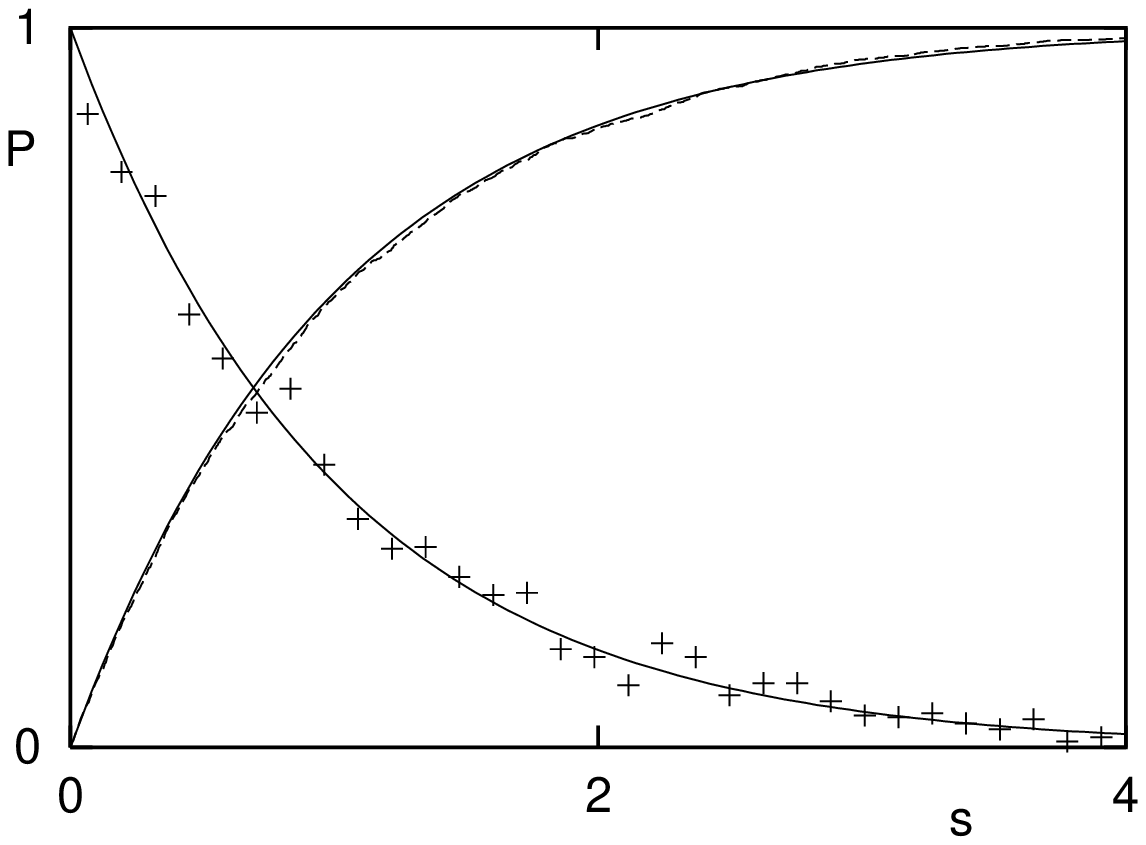} \caption{Level spacing distribution for the symmetry class ${\mathbf G}$.} \label{then:LevelSpacingsPlotG} \end{figure}
\begin{figure} \centering \includegraphics[width=6.8cm,height=11.7cm,angle=-90]{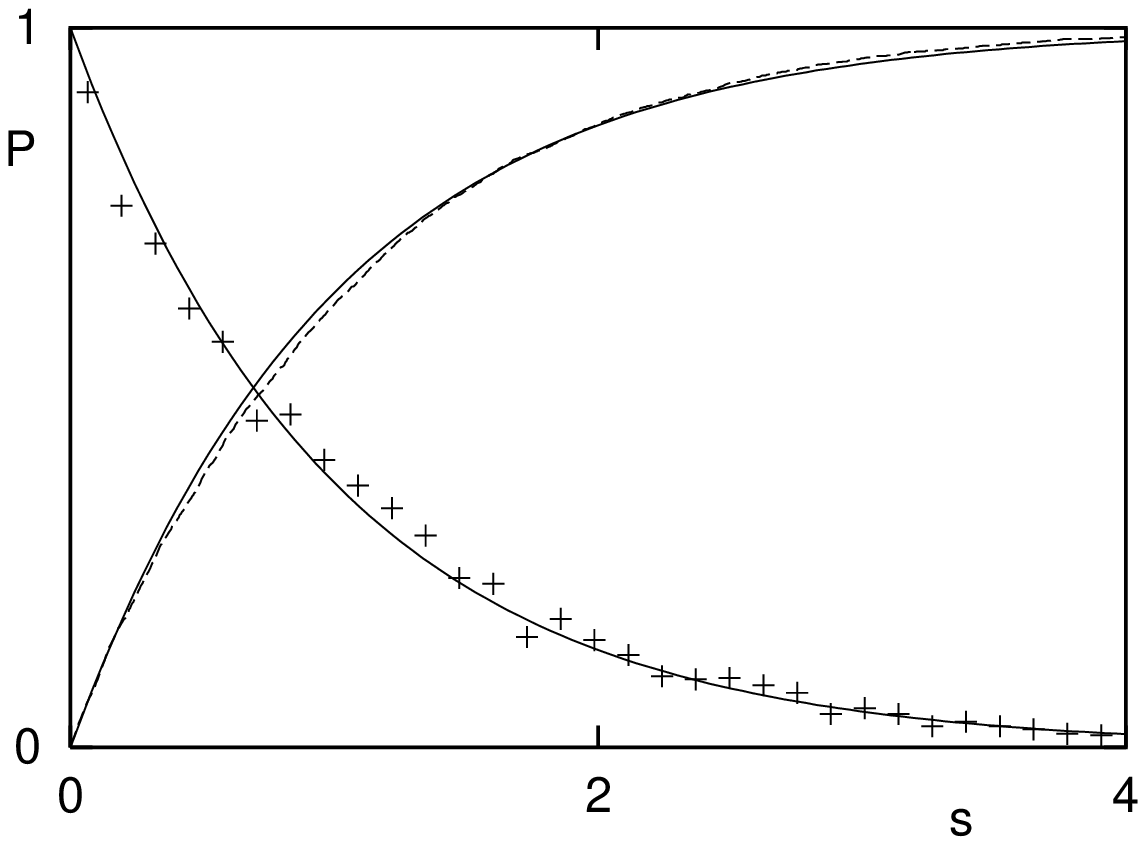} \caption{Level spacing distribution for the symmetry class ${\mathbf C}$.} \label{then:LevelSpacingsPlotC} \end{figure}
\begin{figure} \centering \includegraphics[width=6.8cm,height=11.7cm,angle=-90]{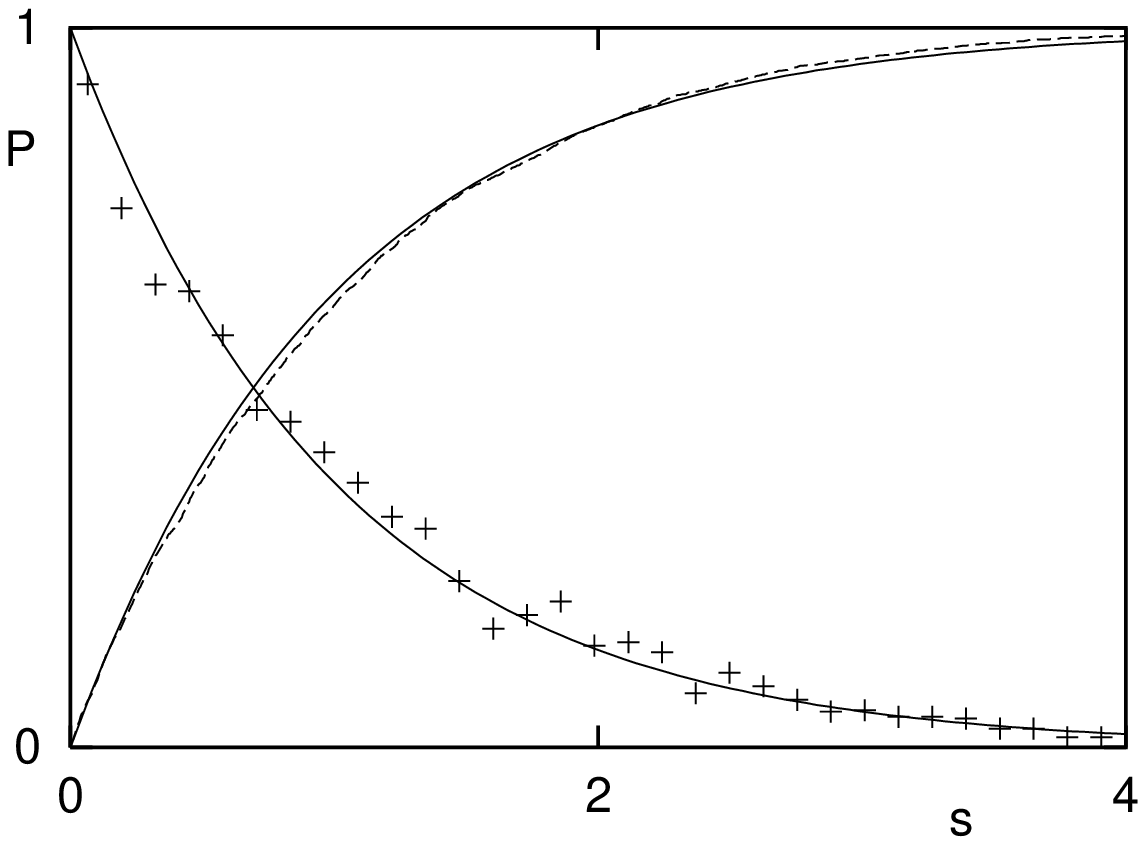} \caption{Level spacing distribution for the symmetry class ${\mathbf H}$.} \label{then:LevelSpacingsPlotH} \end{figure}

Now we are able to regard the spectral fluctuations. We unfold the spectrum,
\begin{align*} x_i=\bar{N}(r_i), \end{align*}
in order to obtain rescaled eigenvalues $x_i$ with a unit mean density. Then
\begin{align*} s_i=x_{i+1}-x_i \end{align*}
defines the sequence of nearest-neighbor level spacings which has a mean value of $1$ as $i\to\infty$. We find that the spacing distribution comes close to that of a Poisson random process,
\begin{align*} P_{\text{Poisson}}(s)=\e^{-s}, \end{align*} see figures \ref{then:LevelSpacingsPlotD} to \ref{then:LevelSpacingsPlotH}, as opposed to that of a Gaussian orthogonal ensemble of random matrix theory,
\begin{align*} P_{\text{GOE}}(s)\simeq \frac{\pi}{2}s\e^{-\frac{\pi}{4}s^2}. \end{align*}
The integrated distribution,
\begin{align*} {\cal I}(s)=\int_{0}^{s}P(t)\,dt, \end{align*}
showing the fraction of spacings up to a given length is also shown in figures \ref{then:LevelSpacingsPlotD} to \ref{then:LevelSpacingsPlotH}.
The spacing distributions of the desymmetrized spectra are in accordance with the conjecture of arithmetic quantum chaos. Also in agreement with the conjecture is that we have not found any degenerate eigenvalues within each symmetry class. But taking the eigenvalues of all four symmetry classes together systematic degeneracies occur due to the following:
\begin{theorem}[Steil \cite{then:Steil1999}] If $\lambda=r^2+1$ is an eigenvalue corresponding to an eigenfunction of the symmetry class ${\mathbf G}$ resp. ${\mathbf H}$ then there exists an eigenfunction of the symmetry class ${\mathbf D}$ resp. ${\mathbf C}$ corresponding to the same eigenvalue. \end{theorem}
These degeneracies were first observed by Huntebrinker \cite{then:Huntebrinker1996} and later explained by Steil \cite{then:Steil1999} with the use of the Hecke operators \cite{then:SmotrovGolovchanskii1991,then:Heitkamp1992}. The Hecke operators are defined by
\begin{align*} \T_{\gamma}g(z)=\frac{1}{|\gamma|}\sum_{\substack{a,b,d\in\mathds{Z}[\ii]-\{0\} \\ ad=\gamma \\ b (\text{mod }d) \\ \Re d>0,\ \Im d\ge0.}}g\big((ad)^{-\frac{1}{2}}(az+b)(d)^{-1}(ad)^{\frac{1}{2}}\big), \quad \gamma\in\mathds{Z}[\ii]-\{0\}. \end{align*}
They are self adjoint operators which commute with the Laplacian and among each other. One can therefore simultaneously diagonalize these operators. The corresponding Maass cusp forms are then called Hecke eigenfunctions. \index{Cusp forms!Maass} The eigenvalue equation of the Hecke operators reads
\begin{align*} \T_{\gamma}g_r(z)=t_{\gamma}g_r(z), \quad \gamma\in\mathds{Z}[\ii]-\{0\}, \end{align*}
where each Hecke eigenfunction is either identical to a Maass cusp form with a given symmetry or to a superposition of Maass cusp forms corresponding to the same eigenvalue $\lambda=r^2+1$, but to different symmetry classes,
\begin{align*} g_r(z)=\sum_{\substack{n\in\mathds{N}\\(\Delta+(r^2+1))f_n(z)=0}}c_nf_n(z). \end{align*}
The Hecke operators are multiplicative,
\begin{align*} \T_{\gamma}\T_{\beta}g_r(z)=\sum_{d|(\gamma,\beta)}\T_{\frac{\gamma\beta}{d^2}}g_r(z), \end{align*}
and the Hecke eigenvalues are connected to the Fourier coefficients,
\begin{align*} b_{\gamma}=b_{1}t_{\gamma}, \quad \gamma\in\mathds{Z}[\ii]-\{0\}, \end{align*}
where the Fourier coefficients $b_{\gamma}$ of the Hecke eigenfunctions are given by
\begin{align*} b_{\gamma}=\sum_{n}c_na_{\gamma,n} \end{align*}
and the index $n$ at the Fourier coefficients of the Maass cusp forms $a_{\gamma}=a_{\gamma,n}$ means that they belong to the Fourier expansion of the $n$-th Maass cusp form $f_n(z)$.
\begin{lemma}[Steil \cite{then:Steil1999}] If $g_r(z)$ is a Hecke eigenfunction that does not vanish identically, then: \par
(i) Its first Fourier coefficient is never zero, $b_1\not=0$. \par
(ii) A Hecke eigenfunction cannot be of symmetry class ${\mathbf G}$ or ${\mathbf H}$. \par
(iii) Hecke eigenfunctions can always be desymmetrized such that they fall either into the symmetry class ${\mathbf D}\cup{\mathbf G}$ or ${\mathbf C}\cup{\mathbf H}$. \end{lemma}
\begin{proof}[Steil's theorem] Let $f_n(z)$ be a Maass cusp form of the symmetry class ${\mathbf G}$ or ${\mathbf H}$. Due to Steil's lemma, it cannot be a Hecke eigenfunction. Since one can diagonalize the Laplacian and the Hecke operators simultaneously, there have to exist linearly independent Maass cusp forms $f_{n+k},\ k=0,\ldots,K$ corresponding to the same eigenvalue $\lambda=r^2+1$ such that
\begin{align*} \sum_{k=0}^{K}c_{n+k}f_{n+k}(z)=g_r(z) \end{align*}
is a Hecke eigenfunction. \par
At least one of these Maass cusp forms has to be of the symmetry class ${\mathbf D}$ or ${\mathbf C}$ in order that
\begin{align*} b_{1}=\sum_{k=0}^{K}c_{n+k}a_{1,n+k} \end{align*}
does not vanish. \par
Since the Hecke eigenfunctions can be desymmetrized such that they fall into the symmetry class ${\mathbf D}\cup{\mathbf G}$ resp. ${\mathbf C}\cup{\mathbf H}$ they are a superposition of either Maass cusp forms of the symmetry classes ${\mathbf D}$ and ${\mathbf G}$ or of Maass cusp forms of the symmetry classes ${\mathbf C}$ and ${\mathbf H}$. Therefore, if $f_n(z)$ is of symmetry class ${\mathbf G}$ one of the $f_{n+k},\ k=1,\ldots,K$ is of symmetry class ${\mathbf D}$, and if $f_n(z)$ is of symmetry class ${\mathbf H}$ one of the $f_{n+k},\ k=1,\ldots,K$ is of symmetry class ${\mathbf C}$. \end{proof}
Based on our numerical results we now conjecture the following:
\begin{conjecture} Taking all four symmetry classes together, there are no degenerate eigenvalues other than those explained by Steil's theorem. Furthermore, the degenerate eigenvalues which are explained by Steil's theorem occur only in pairs of two degenerate eigenvalues. They never occur in pairs of three or more degenerate eigenvalues. \end{conjecture}
Maass cusp forms of the symmetry classes ${\mathbf G}$ and ${\mathbf H}$ indeed occur. On the one hand we have found a number of them numerically. On the other hand, Weyl's law also explains their existence. Due to this the number of eigenvalues whose corresponding Maass cusp forms belong to a specific symmetry class is in leading order independent of the choice of the symmetry class. Weyl's law together with Steil's theorem lead to the following:
\begin{conjecture} The sequence of non-degenerate eigenvalues in the spectrum of the Laplacian for the Picard group is of density zero. \end{conjecture}
This means that as $\lambda\to\infty$
\begin{align*} \frac{\#\{\text{non-degenerate eigenvalues}\le\lambda\}}{\#\{\text{degenerate eigenvalues}\le\lambda\}}\to0. \end{align*}
Table \ref{then:FirstFewEigenvalues} looks as if it would contradict this conjecture. But this is due to the fact that only the first few eigenvalues are listed. In table \ref{then:SomeLargeEigenvalues} we list some consecutive large eigenvalues where we can see a better agreement with the conjecture.
\begin{table} \caption{Some consecutive large eigenvalues of the Laplacian for the Picard group. Listed is $r$, related to the eigenvalues via $\lambda=r^2+1$.} \label{then:SomeLargeEigenvalues} \begin{align*} {\mathbf D}& & {\mathbf G}& & {\mathbf C}& & {\mathbf H}& \\ \\
139&.65419675 & 139&.65419675 & 139&.66399548 & 139&.66399548 \\
139&.65434417 & 139&.65434417 & 139&.66785333 & 139&.66785333 \\
139&.65783548 & 139&.65783548 & 139&.66922266 & 139&.66922266 \\
139&.66104047 & 139&.66104047 & 139&.67870460 & 139&.67870460 \\
139&.67694018 &    &          & 139&.68234200 & 139&.68234200 \\
139&.68162707 & 139&.68162707 & 139&.68424704 & 139&.68424704 \\
139&.68657976 &    &          & 139&.69369972 & 139&.69369972 \\
139&.71803029 & 139&.71803029 & 139&.69413379 & 139&.69413379 \\
139&.72166907 & 139&.72166906 & 139&.69657741 & 139&.69657741 \\
139&.78322452 & 139&.78322452 & 139&.73723373 & 139&.73723373 \\
139&.81928622 & 139&.81928622 & 139&.73828541 & 139&.73828541 \\
139&.81985670 & 139&.81985670 & 139&.74467774 & 139&.74467774 \\
139&.82826034 & 139&.82826034 & 139&.75178180 & 139&.75178180 \\
139&.84250751 &    &          & 139&.75260292 & 139&.75260292 \\
139&.87781072 & 139&.87781072 & 139&.79620628 & 139&.79620628 \\
139&.87805540 &    &          & 139&.80138072 & 139&.80138072 \\
139&.88211647 & 139&.88211647 & 139&.81243991 & 139&.81243991 \\
139&.91782003 & 139&.91782003 & 139&.81312982 & 139&.81312982 \\
139&.91893517 &    &          & 139&.82871870 & 139&.82871870 \\
139&.92397167 & 139&.92397167 & 139&.86401372 & 139&.86401372 \\
139&.92721861 & 139&.92721861 & 139&.86461581 & 139&.86461581 \\
139&.93117207 & 139&.93117207 & 139&.89407865 & 139&.89407865 \\
139&.93149277 & 139&.93149277 & 139&.89914777 & 139&.89914777 \\
139&.94067283 &    &          & 139&.90090849 & 139&.90090849 \\
139&.94396890 & 139&.94396890 & 139&.91635302 & 139&.91635302 \\
139&.95074070 &    &          & 139&.94071729 & 139&.94071729 \\
139&.95124805 & 139&.95124805 & 139&.95080198 & 139&.95080198 \\
139&.99098324 & 139&.99098324 & 139&.97043676 & 139&.97043676 \\
\end{align*} \end{table}

\section{Summary} Our principal goal was to test the conjecture of arithmetic quantum chaos numerically in one example. For this purpose we have chosen a point particle moving freely in the three-dimensional and negatively curved quotient space of the Picard group. Identifying the solutions of the stationary Schr\"{o}dinger equation with Maass waveforms allowed us to use Hejhal's algorithm to compute the eigenfunctions and eigenvalues numerically. Having computed $13950$ eigenvalues (and eigenfunctions), which exceeds all previous computations in non-integrable three-dimensional systems, we demonstrated that our numerical results are in accordance with the conjecture of arithmetic quantum chaos. Within each symmetry class we do not find any degenerate eigenvalues, but taking all four symmetry classes together, almost all eigenvalues become degenerate in the limit of large eigenvalues $\lambda\to\infty$. This behaviour was explained by the interplay of the symmetries with the Hecke-operators.

\section{Acknowledgments} The help of Ralf Aurich, Jens Bolte, Dennis A. Hejhal and Frank Steiner is gratefully acknowledged. The author is supported by the Deutsche Forschungsgemeinschaft under the contract no. DFG Ste 241/16-1. Part of the work was done while I was a member of Dennis A. Hejhal's group in Uppsala (Sweden) supported by the European Commission Research Training Network HPRN-CT-2000-00103. The computations were run on the Universit\"{a}ts-Rechenzentrum Ulm.

\appendix

\section{The K-Bessel function} \label{then:KBessel} The K-Bessel function is defined by
\begin{align*} K_{\ii r}(x)=\int_{0}^{\infty} \e^{-x\cosh t} \cos(r t) \, dt, \quad \Re x>0, \quad  r\in\mathds{C}, \end{align*}
see Watson \cite{then:Watson1944}, and is real for real arguments $x$ and real or imaginary order $\ii r$. It solves the modified Bessel differential equation
\begin{align*} x^2 u''(x) + x u'(x) - (x^2 - r^2) u(x) = 0, \end{align*}
and decays exponentially for large arguments
\begin{align} \label{then:LargeArgumentK} K_{\ii r}(x)\sim\sqrt{\frac{\pi}{2x}} \e^{-x} \quad \text{for } x\to\infty. \end{align}
A second linearly independent solution of the modified Bessel differential equation is the I-Bessel function
\begin{align*} I_{\ii r}(x)=(\frac{x}{2})^{\ii r} \sum_{k=0}^{\infty} \frac{(\frac{x}{2})^{2k}}{k!\Gamma(\ii r+k+1)}, \end{align*}
which grows exponentially for large arguments
\begin{align*} I_{\ii r}(x)\sim\sqrt{\frac{1}{2\pi x}} \e^{x} \quad \text{for } x\to\infty. \end{align*}
The K-Bessel function decreases exponentially when $r$ increases. This can be compensated by multiplication with the factor $\e^{\frac{\pi r}{2}}$. \par
In order to compute the K-Bessel function numerically for small or moderate imaginary order we use its continued fraction representation which follows from the Miller algorithm \cite{then:Temme1975}.


\end{document}